\documentstyle[aps,prc,epsf]{revtex}

\begin{document}


\title{Pauli exclusion operator and binding energy of nuclear matter}

\author{E.~Schiller, H.~M\"uther}
\address{Institut f\"ur Theoretische Physik, Universit\"at T\"ubingen,
         D-72076 T\"ubingen, Germany}
\author{P.~Czerski}
\address{Instytut Fizyki J\c adrowej, Pl-31-342 Krak\'ow, Poland}

\maketitle

\begin{abstract}
Brueckner-Hartree-Fock calculations are performed for nuclear matter with an
exact treatment of the Pauli exclusion operator in the Bethe-Goldstone equation.
The differences in the calculated binding energy, compared to the angle-average 
approximation, which is commonly used, are non-negligible. These difference
exhibits a specific density dependence, which shifts the calculated saturation
point towards smaller densities. This effect is observed for various versions of
modern models for the NN interaction. 
\end{abstract}


It is one of the very central and very old projects of nuclear structure theory,
to evaluate the saturation properties of nuclear matter from a realistic
nucleon-nucleon (NN) interaction without any adjustment of a free parameter. The
so-called hole-line expansion or Brueckner-Bethe-Goldstone theory has been one
of the tools for solving this many-body problem, which has already been used for
many years\cite{brueck,bethe,day1,haftel}. These early investigations were
successful to some extent. The inclusion of NN correlations in the lowest order
of the hole-line expansion, the Brueckner-Hartree-Fock (BHF) approach, turned
out to be very important. Realistic models of the NN interaction like the Reid
soft-core potential\cite{reid} yield an energy of nuclear matter around 150 MeV
per nucleon if the effects of correlations are ignored in a mean-field or
Hartree-Fock calculation. The BHF approach provides a drastic improvement
leading to an energy per nucleon of $-11$ MeV, which was only by 5 MeV off from 
the empirical value of $-16$ MeV per nucleon. Attempts have been made to improve
the description of the saturation point further by exploring different NN
interactions. It turned out, however, that BHF calculations using these various
NN interactions yield results for the saturation point, which fall on the
so-called Coester band\cite{coester}. They either predict too small binding
energy at the empirical value for the density, or about the correct energy at a
density, which is too large by a factor of two, or results in between.
Comparison of BHF  with variational calculations furthermore demonstrated that
the inclusion of three-hole line contributions seems to be necessary to obtain a
reliable estimate for the binding energy of nuclear matter\cite{vari,day2}.

During the last years some progress has been made in this field. It has been
shown that a continuous choice for the particle spectrum\cite{baldo,mahaux} (see
also discussion below) accounts for the main part of the effects of three-body 
correlations. The discrepancy between the calculated saturation points of
nuclear matter and the empirical one has significantly been reduced by
considering relativistic effects within the Dirac-Brueckner-Hartree-Fock (DBHF)
approach\cite{shakin,brock,malf}. Finally, it should be mentioned that a new
generation of realistic NN potentials has been developed\cite{nim,v18,cdbonn},
which yield very accurate fits of proton-neutron and proton-proton scattering.
These new potentials, which are essentially phase-shift equivalent remove a
large part of the discrepancies observed between older models of the NN
interaction\cite{mort}.

Because of these improvements, the time seems to be appropriate to test 
the reliability of approximations which are generally employed in the BHF
calculations of nuclear matter. One of the central equations to be solved in the
BHF approximation is the Bethe-Goldstone equation, which we may write in the
momentum representation
\begin{equation}
G_{S,T}(\vec k,\vec k', K,\omega ) = V_{S,T}(\vec k, \vec k') + \int
d^3p\, V_{S,T}(\vec k, \vec p) \frac{Q(\vec p,K)}{\omega - H_0}
G_{S,T}(\vec p,\vec k', K,\omega ) \label{eq:bg}
\end{equation}
In this equation $K$ represents the center of mass momentum of the interacting
pair of nucleons while $\vec k$, $\vec k'$ and $\vec p$ stand for relative
momenta, which are related to the single-particle momenta according to
\begin{equation}
\vec K = \frac{1}{2} \left( \vec k_1 + \vec k_2 \right) \quad
\mbox{and}\quad 
\vec k = \frac{1}{2} \left( \vec k_2 - \vec k_1 \right)\, ,
\end{equation}
$\omega$ denotes the starting energy, while $V_{S,T}$ and $G_{S,T}$ refer to 
the matrix elements of the bare interaction and $G$-matrix respectively for two
nucleons with total spin $S$ and isospin $T$. The
operator $H_0$ is used to define the energy spectrum of the intermediate
two-particle state ($K,\vec p$). The conventional choice for these energies of
particle states above the Fermi surface has been to replace $H_0$
by the kinetic energy of these states. In the so-called
continuous choice, one assumes that the single-particle energies for these
particle states as well as for the hole states are calculated from the kinetic
energy plus a mean field contribution, which is calculated in a self-consistent
way from the $G$-matrix by
\begin{eqnarray}
\epsilon_q  & = &\frac{q^2}{2m} + \int_{p\leq k_F} d^3p \, \langle \vec q \vec p \vert
G(\omega = \epsilon_q + \epsilon_p) \vert\vec q \vec p \rangle \nonumber \\
& \approx & \frac{q^2}{2m^*} + U \, .\label{eq:epsi}
\end{eqnarray}
The second line exhibits a parametrisation of these single-particle energies
in terms of an effective mass and a constant potential, which is often used.
The Pauli operator in the Bethe-Goldstone equation (\ref{eq:bg}) prevents
scattering into intermediate states with momenta $\vec p_1 = \vec K - \vec p$
and $\vec p_2 = \vec K + \vec p$, which are smaller than the Fermi momentum
$k_F$. Therefore the value of $Q(\vec p,K)$ depends on the angle $\Omega$
between the center of mass
momentum $K$ and the relative momentum $p$. Employing a partial wave expansion
of the two-particle states, the matrix elements for this Pauli operator can be
written\cite{cheon}
\begin{eqnarray}
\langle (l'S)J'M\vert Q(p, K) \vert (lS)JM\rangle & = & \sum_{m_l, m_S} 
(l'm_lSM_s \vert J'M)(JM\vert lm_lSM_s) \nonumber \\
&& \qquad \times \langle l'm_l\vert Q(p, K) \vert lm_l\rangle\quad
\mbox{with}\label{eq:paul}\\
\langle l'm_l\vert Q(p, K) \vert lm_l\rangle & = & \int d\Omega \, Y^*_{l'm_l} (\Omega )   
Y_{lm_l} (\Omega ) \Theta \left( \left| \vec K + \vec p \right| -k_F\right) 
\Theta \left( \left| \vec K - \vec p \right| -k_F\right)\,, \nonumber
\end{eqnarray}
with $\Theta (x)$ defining the step function.
One finds that the Pauli operator is diagonal with respect to the modulus of
the momenta $K$ and $p$, the spin $S$ and the projection quantum number $m$. It
has non vanishing matrix elements between states of different $l$ and $J$ with the
restriction of parity conservation ($l+l'$ must be even) and its value depends
on the projection quantum number $M$. This implies that also
the solution of the Bethe-Goldstone equation is not diagonal with respect to the
angular momentum quantum numbers $l$ and $J$ and depends on $M$,
\begin{equation}
\langle klJ\vert G_{STM} (\omega, K) \vert k'l'J'\rangle \, . \label{eq:gexact}
\end{equation} 
In order to simplify the calculation this Pauli operator is usually replaced 
by the so-called angle-average approximation
\begin{equation}
\langle l'm_l\vert Q^A(p, K) \vert lm_l\rangle = \delta_{l,l'}\cases{0 & for
$p\leq\sqrt{k_f^2-K^2}$, \cr 1 & for $p\geq k_F+K$, \cr
\frac{K^2+p^2-k_F^2}{2Kp} & else\cr}
\label{eq:qaa}
\end{equation}
The angle-average approximation yields matrix elements for the Pauli operator
which are diagonal in the $l$ and $J$ and independent on $M$. This means that
the Bethe-Goldstone eq.~(\ref{eq:bg}) can be solved separately for each partial
wave and the resulting $G$-matrix will be diagonal in $J$ and independent on
$M$. 

The matrix elements of $G$ can then be used to evaluate the total energy per
nucleon. In the case of the exact Pauli operator this energy is given as
\begin{eqnarray}
\frac{E}{A} & = & \frac{3}{5}\frac{k_F^2}{2m}\quad  + \quad \frac{6}{ k_F^3}
\sum_{\stackrel{T,S,M,l',l,}{J',J,m_l,m_s}} (2T+1) \int k^2 dk
\int K^2 dK \int d\Omega  \nonumber \\
&& \qquad \times \langle klJ\vert G_{STM} (\omega, K) \vert k l'J'\rangle 
(l'm_lSm_S\vert J'M) 
(lm_lSm_S\vert JM) Y_{l'm_l} (\Omega) Y_{lm_l} (\Omega ) \nonumber \\
&&\qquad \quad \times\Theta (k_F - \vert
\vec K + \vec k \vert ) \Theta (k_F - \vert\vec K - \vec k \vert ) 
\label{eq:eexact}
\end{eqnarray}
with $\Omega$ the angle between the direction of the relative momentum $\vec k$
and the center of mass momentum $\vec K$. In the calculations discussed below we
consider the coupling of partial waves up to $J,J'\leq 6$. For angular momenta
larger than 6, the Born approximation is used. If the matrix elements of $G$ are
diagonal in the total angular momentum ($J=J'$) and independent on the projection
quantum number $M$, as it is the case in the angle-average approximation for the
Pauli operator, this expression can be rewritten into
\begin{eqnarray}
 \frac{E}{A} & = & \frac{3}{5}\frac{k_F^2}{2m} + \frac{6}{ k_F^3}
\sum_{T,S,l,J} (2T+1)(2J+1) \int_0^{k_F} dk\, k^2 \Biggl[ \int_0^{k_F-k} dK\, K^2
\nonumber \\ &&\qquad + \int_{k_F-k}^{\sqrt{k_F^2-k^2}} dK\, K^2 \frac{k_F^2 -
K^2 -k^2}{2Kk}\Biggr] \langle klJ \vert G_{ST}(\omega, K) \vert klJ\rangle\, .\label{eq:eapro}  
\end{eqnarray}
which corresponds to the standard expression discussed e.g.~in \cite{haftel}.

\begin{figure}[t]
\epsfysize=10.0cm
\begin{center}
\makebox[16.4cm][c]{\epsfbox{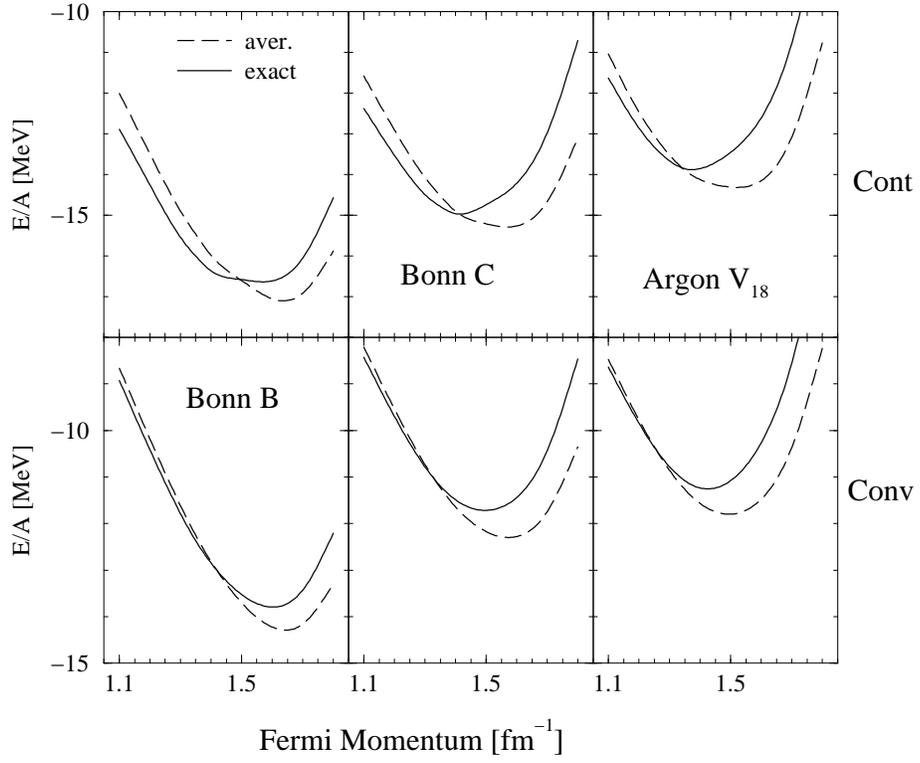}}
\end{center}
\caption{Calculated binding energies per nucleon for nuclear matter as a
function of the Fermi momentum $k_F$. Results are presented for the
angle-average approximation of the Pauli operator (dashed lines) and the exact
treatment (solid lines). The potentials Bonn B and C, defined in 
\protect\cite{rupr},
as well as the neutron-proton part of the Argonne
$V_{18}$ potential \protect\cite{v18} were used for the NN interaction. The
continuous choice for the particle state spectrum in the Bethe-Goldstone eq.(\ref{eq:bg}) has
been used to obtain the results displayed in the upper part, while the
conventional choice has been used to calculate the results shown below.}

\label{fig:one}

\end{figure}

Results for the calculated binding energy per nucleon as a function of the Fermi
momentum are displayed in Fig.~\ref{fig:one}, considering various NN potentials.  
The continuous choice (figures in the upper half) as well as the conventional
choice were used for the single-particle spectrum. For each density, NN
interaction and particle state spectrum, a self consistent single-particle
spectrum has been determined using the effective mass parametrisation defined in
eq.~(\ref{eq:epsi}). The total binding energy has then been calculated using 
the angle-average approximation for the Pauli operator (dashed lines)
as well as the exact treatment (solid lines). 

A general feature can be observed, which is independent on the NN interaction 
and the choice for the single-particle spectrum: The angle-average approximation
tends to underestimate the binding energy per nucleon at low densities but
overestimates it at higher densities. The effects of the exact treatment of the
Pauli operator is not very large around and below the empirical value for the saturation 
density. This is in agreement with older studies of the angle-averaged
approximation\cite{cheon,leng}, in which matrix elements of $G$ were compared at
those small densities. The characteristic density dependence for the exact
treatment of the Pauli operator, however, leads to a non-negligible shift in the
calculated saturation, i.e.~the minimum in the energy versus density curve.
This correction moves the calculated saturation points to smaller densities and
smaller energies. It is worth noting that the saturation points calculated for
the Bonn C potential (E$=-14.97$ MeV at $k_F$=1.42 fm$^{-1}$) and the
neutron-proton part of Argonne $V_{18}$ potential (E$=-13.89$ MeV at $k_F$=1.37
fm$^{-1}$) are rather close to the empirical value.

The effects of  an exact treatment of the Pauli operator are larger for the 
continuous choice of the single-particle spectrum than for the conventional one. 
This is quite plausible as the continuous choice yields a larger sensitivity to the
proper treatment of states around the Fermi momentum. A similar argument can be
used to explain the fact that the Pauli effects are larger for NN potentials
which have a slightly stronger tensor force (Bonn C and Argonne $V_{18}$) than is
observed for the Bonn B potential, which contains a weaker tensor component.

In summary we would like to point out that the effects of an exact treatment of
the Pauli operator in the Bethe-Goldstone equation are not very dramatic in
particular for nuclear matter at small densities. A characteristic density
dependence of these Pauli corrections, however, lead to a non-negligible
improvement in the calculated saturation points. Therefore many-body
calculations going beyond the BHF approximation should take these effects into
account.

This work has been supported  by the Graduiertenkolleg `Struktur und
Wechselwirkung von Hadronen und Kernen' (DFG, GRK 132/3) and the scientific
exchange program between Germany and Poland (POL-246-96).

\end{document}